\documentclass[conference]{IEEEtran}
\IEEEoverridecommandlockouts
\usepackage{cite}
\usepackage{amsmath,amssymb,amsfonts}
\usepackage{algorithmic}
\usepackage{graphicx}
\usepackage{moresize}
\usepackage{textcomp}
\usepackage[T1]{fontenc}
\usepackage{lmodern}
\usepackage[utf8x]{inputenc}
\usepackage[
  separate-uncertainty = true,
  multi-part-units = repeat
]{siunitx}
\usepackage{xcolor}
\def\BibTeX{{\rm B\kern-.05em{\sc i\kern-.025em b}\kern-.08em
    T\kern-.1667em\lower.7ex\hbox{E}\kern-.125emX}}

\begin{document}

\title{Low-Cost Device Prototype for Automatic Medical Diagnosis Using Deep Learning Methods
}

\author{\IEEEauthorblockN{Neil Deshmukh}
\IEEEauthorblockA{{Moravian Academy} \\
Macungie, USA \\
neil.nitin.de@gmail.com}

}

\maketitle

\begin{abstract}
This paper introduces a novel low-cost device prototype for the automatic diagnosis of diseases, utilizing inputted symptoms and personal background. The engineering goal is to solve the problem of limited healthcare access with a single device. Diagnosing diseases automatically is an immense challenge, owing to their variable properties and symptoms. On the other hand, Neural Networks have developed into a powerful tool in the field of machine learning, one that is showing to be extremely promising at computing diagnosis even with inconsistent variables. 

In this research, a cheap device (under \$30) was created to allow for straightforward diagnosis and treatment of human diseases. By utilizing Deep Neural Networks (DNNs) and Convolutional Neural Networks (CNNs), outfitted on a Raspberry Pi Zero processor (\$5), the device is able to detect up to 1537 different diseases and conditions and utilize a CNN for on-device visual diagnostics. The user can input the symptoms using the buttons on the device and can take pictures using the same mechanism. The algorithm processes inputted symptoms, providing diagnosis and possible treatment options for common conditions. The purpose of this work was to be able to diagnose diseases through an affordable processor with high accuracy, as it is currently achieving an accuracy of 90\% (\textpm 0.8\%) for Top-5 symptom-based diagnoses, and 91\% (\textpm 0.2\%) for visual skin diseases. The NNs achieve performance far above any other tested system (WebMD, MEDoctor, so forth.), and its efficiency and ease of use will prove it to be a helpful tool for people around the world. This device could potentially provide low-cost universal access to vital diagnostics and treatment options.

\end{abstract}
\section{Introduction}
Communicable diseases, maternal causes, and nutritional deficiencies, called Group I conditions, cause more than half of all deaths in low income countries \cite{b1}. More than 400 million people do not have access to essential health services, with an average of one doctor for every 2000 people in developing countries \cite{b2}. Many of these people struggle to get enough food and thrive, so many do not have the time nor money to go to a distant hospital. In regions of Africa and India, there is a single doctor for every 20,000 people\cite{b3} (see Fig. 1). Even more concerning is that these doctors usually centralize in urban areas, leaving rural villages with little or no medical support\cite{b4}. Many within range of medical facilities still are not provided attention, as even the care is unable to keep up with the hundreds of people that require aid, resulting in further death\cite{b5}. In low-income populations, the morbidity of diseases increases, due to overcrowding and increased transmission, and outbreaks of disease are more frequent and more severe when the population density is high. Such examples of these diseases with epidemic potential that have high transmission rates are: acute respiratory infections, meningitis, typhus, cholera, and scabies\cite{b6}. Up to half of the world's population are at risk of many endemic diseases. 50,000 people are dying every day from infectious diseases; in actuality, these diseases are easily treatable, but, unfortunately, many people do not have the treatment knowledge\cite{b7}. In addition, millions of people are developing cancers as a direct result of preventable infections by bacteria and viruses\cite{b8}. 
\begin{figure}[htbp]
\centerline{\includegraphics{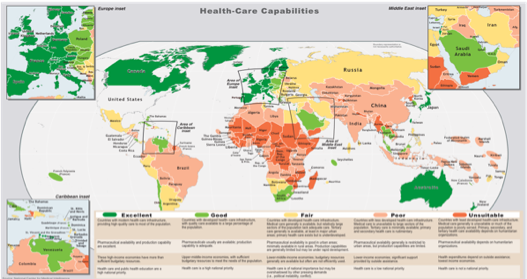}}
\caption{Map of Healthcare Coverage \cite{b15}}
\label{fig}
\end{figure}
About 80 percent of people in developing countries rely on local health care for medical assistance. This is often in the form of ill-staffed clinics and local healers\cite{b5}. There have been many clinics set up worldwide, but these still only have limited reach, and it takes an amount of time to instill recognition skills among primary care workers\cite{b9}. Many people still are unreachable by our means and suffer through their daily life with diseases, ones that are easily treatable. Many, in fact, have access to the treatment necessary to prevent further detriment, but due to the limited nature of healthcare in developing countries, they do not have access to the knowledge\cite{b7}. Adding to this significant problem, 17\% of the population was tipped or pushed further into extreme poverty (\$1.25/day) because they had to pay for health services out of their own pockets. The average cost of a medical checkup is around \$15-\$20 for a thorough checkup, which is far above the price point for many people in desperate need of medical attention\cite{b2}. The unfortunate fact is that more than one billion people, one out of every seven, live in extreme poverty (\$1.25/day), and 2.2 billion people live on less than \$2 a day\cite{b10}. These clinics might be too far away to tempt any locals to travel for a simple diagnosis, and instead, they might have a chance of dying with an easily treatable disease. This problem is not only isolated in developing areas but also in developed countries, like the US, as well. In the US alone, 45,000 people die every year, due to a dismissal of their symptoms and their refusal to go to a healthcare facility\cite{b11}. A single checkup costs upwards of \$50, which deters those to immediately following up if noticing a detriment or symptom\cite{b12}. 

The tools that they are given are also incredibly inaccurate, as the most popular symptom checkers have an abysmal average diagnosis accuracy of 34\%\cite{b13}.  These symptom checker’s utilize a database to match symptoms to diseases, overlooking the complex and interwoven relationship between symptoms and diseases. A few ML based diagnostic options are available, but they are all highly privatized, and some are developed to be used in hospitals, rather than by individual people. All current ML based diagnostic solutions also rely on cloud based models, making diagnosis without a stable internet connection impossible. The only other available similar research was carried out by a team at Stanford University to develop an algorithm for visual diagnosis of skin diseases. The team transfer-trained a pretrained Inceptionv3 network on images of skin diseases\cite{b19}. They obtained an accuracy of 71.9\% with their algorithm using a multinodal approach. The pretrained model was trained on images of everyday objects, so the features detected might not have corresponded to skin diseases well, as compared to an algorithm trained from scratch.

Developing NNs to solve both of these problems through a singular device has the potential to revolutionize healthcare, and help people without access to medical facilities.

\section{Methods}

\subsection{Development of DNN}

Diseases have variability in their expression of certain symptoms, which makes it difficult to determine and classify the specific disease of the user. Past attempts include traditional symptom checkers, which extract diagnoses through a branching symptom structure, failing if one symptom node is not expressed. Other algorithms require large amounts of memory, usually on servers, to run models that are more elaborate versions of the branching structure, simply with more inputs. In order to solve this diagnosis dilemma, a DNN approach was utilized, with a multilayered network structure with symptoms as the nodes in the model (see Fig. 2). A DNN is able to learn and model non-linear and complex relationships, which are the case for medical diagnoses. Most importantly, after learning from the initial inputs and their relationships, it can infer relationships on unseen data as well, making the model generalize and predict on unseen data, allowing the DNN to account for real-world variables through symptoms and diagnosis. The Tensorflow Machine Learning Library\cite{b26}, was used to develop, train, test, and optimize the NNs. The DNN was trained on inputs of 237 possible symptoms relating towards 1537 diagnoses, with a structure containing 18 layers (15\% dropout optimization for every other layer) and 11,546,629 trainable parameters. The model does not require hand-crafted features coded directly to diagnose, but instead was trained, through solely inputs and end diagnoses. The model takes as many possible inputs as the user gives, providing five possible diagnoses with treatments.
\begin{figure}[htbp]
\centerline{\includegraphics{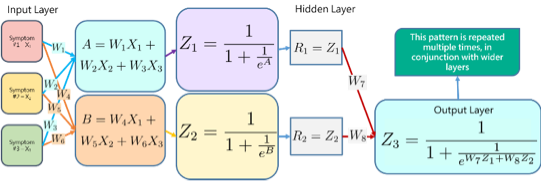}}
\caption{Simplified Structure of DNN \& Operations Between Layers}
\label{fig}
\end{figure}
\subsection{Development of CNN}
To increase the diagnostic capabilities of the device, another neural network was constructed to utilize the camera to diagnose diseases on the epidermis. Compared to the 48 layers and 7 million parameters from Inceptionv3\cite{b23}, utilized by the Stanford research team\cite{b19}, the Inception-Resnetv2\cite{b22} CNN structure has 467 layers with more than 54 million parameters. This CNN consists of a relatively new method in machine learning called a  residual module\cite{b27}. In theory, the deeper the network is, or the more layers it consists of, the more accurate it should become. In reality, however, the increased depth results in the signal to change the weights becoming very small in the early layers, resulting in these layers becoming negligible learned, or called a vanishing gradient. The second problem with training deeper networks is, performing the optimization on a huge parameter space and therefore naively adding the layers, leads to higher training, and therefore testing, errors, which is called degradation. To solve this, the residual module creates a direct path between the input and output to the module implying identity mapping and the added layer needs to learn the features on top of the already available input. The network also consists of convolution, pooling, concat, dropout, fully connected, and softmax layers (see Fig. 3). 
\begin{figure}[htbp]
\centerline{\includegraphics{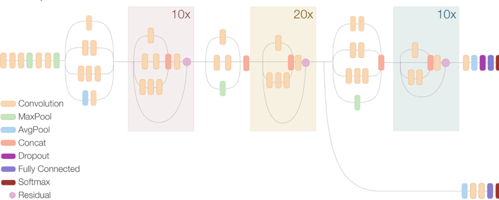}}
\caption{Structure of CNN \cite{b16}}
\label{fig}
\end{figure}
\subsection{Training of Neural Networks}
After the NNs had been constructed, each one was trained from scratch individually, solely on their raw inputs and outputs. The inputs for the disease DNN were raw symptoms, corresponding to one-hot encoded values in the created database, and the diagnoses as one-hot encoded outputs.  This dataset comes from the open-source repository at PubMed\cite{b20}, with 237 symptoms relating towards 1537 different diagnoses. The dataset contained 849,103 records from PubMed, which were used to train without distortion. The DNN was trained over the course of 1000 epochs, back and forth computations of the model, using a ReLU function and a learning rate of 0.001, and utilizing an ADAM optimizer.

For the CNN, the dataset consisted of 26 skin afflictions, with 3,050,784 images, after image preprocessing and database manipulation. These were trained across a CNN, for the skin afflictions. These images were extracted through the open-source databases at DermNet\cite{b21}. These manipulations included rotation across a 30-degree basis, a Gaussian blur through OpenCV\cite{b24}, a white noise filter, and an increase or decrease in the total brightness of the images. The dataset was also split into 2,986,121 training and validation images and 30162 testing images. These repositories were then frozen into TFRecord files to speed up training. The hyperparameters used were utilized through backpropagation, with, over the course of 500 epochs, a batch size of 256, initial learning rate of 0.0007, a learning rate decay factor of 0.7, and 2 epochs before decay. The first model was trained with a weight decay of 0.00004, a batch norm decay of 0.9997, a batch norm epsilon of 0.001, utilizing an ADAM optimizer, and with ReLU as the activation function for Conv2D. For the dropout layer, there was an 15\% drop probability. These hyperparameters were summarized after a series of tests using different parameters. All of the previously mentioned training was conducted on a research computer, utilizing 32 GBs of GPUs, 8 vCPUs and, 64 GBs of RAM.

\subsection{Development of Prototype: Low-Cost Device to Run NNs}
The principal objectives of the prototype were centralized around accuracy, affordability, and usability. Implementing a Raspberry Pi, a \$5 processor, proved to identify these problems. Unfortunately, the meager amount of processing power on the Pi is unable to normally process these NNs, so certain precautions must be taken to allow them to run on the Pi. The outputs of the training process were stored as checkpoint files, which were frozen to decrease their size into a single file. TensorFlow GraphDefs, produced by the training code, contain all the computation that is needed for back-propagation and updates of weights, as well as the queuing and decoding of inputs and the saving out of checkpoints \cite{b18}. None of these artifact nodes from training are needed during inference, so they were removed through pruning. In order to decrease the Floating Point Operations per Second (FLOPS) computation necessary for the Pi, through a process called quantization, the constant parameters were rounded from 32-bit floats to 8-bit floats. The primary obstacle comes in the form of the memory and processing needed to run the model. A Raspberry Pi Zero can compute around 3 billion FLOPs while the models described around 15 billion FLOPS. The NNs were allowed to run on the Pi through memory mapping, where a single file in which the first part is a normal GraphDef serialized into the protocol buffer wire format, but then the weights are appended in a form that can be directly mapped. The original size of the NN was 210 MB, but after the quantization techniques utilized, it reduced to 53 MB, with only a negligible decrease in accuracy. Memory mapping sped up the runtime of the CNN on the Raspberry Pi by 44\% (optimized runtime of CNN:14s). The runtime of the DNN for symptom analysis is 6s.

\begin{figure}[htbp]
\centerline{\includegraphics[scale=1.1]{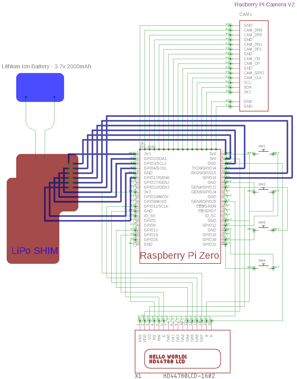}}
\caption{Circuit Diagram of Device}
\label{fig}
\end{figure}
The Neural Networks were then implemented into a User Interface, solely with the input symptoms and modes being inputted through buttons, and displayed on the LCD screen (see Fig. 4 \& 5).  The final layer of the graph computes the probabilities of the classes. The most affordable version of the prototypes costs than \$15 to diagnose diseases, entirely offline, but still accurately.
\begin{figure}[htbp]
\centerline{\includegraphics{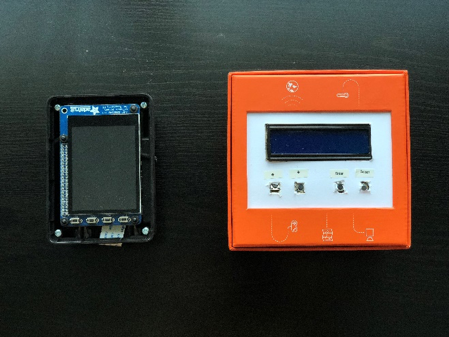}}
\caption{Pictures of Prototypes (Left: touchscreen based | Right: button based)}
\label{fig}
\end{figure}

\section{Data Validation}
Due to the variable nature of real-world medical problems, the validity of the NNs was assessed using multiple techniques. The outputs of each checkpoint were evaluated through TensorBoard, compiling histograms and graphs of loss curves and activations. The performance of the models were measured using two methods, cross-entropy, and pure accuracy. The cross entropy loss function was used to measure the measure of backpropagation for all the NNs. To conclusively validate the efficiency of the algorithms, on-device testing was used to obtain the values described. The disease DNN outputs probability values for each disease, which were utilized in the analysis. The DNN was also validated using two different metrics: top one accuracy, the percentage of cases where the correct diagnosis was the prediction that the DNN was most sure with; and top five accuracy, the percentage of cases where the correct disease was in the top five predicted diseases by the DNN. It achieved around a 61\% (\textpm 1.3\%) top one accuracy, and a 90\% (\textpm 0.8\%) top five accuracy. This is an enormous leap from the abysmal 35\% accuracy rate of WebMD and similar non-NN symptom checkers. In order to understand the connections between the diseases, a map was created in dimensional space to visualize the relationships between certain diseases and images. In the case of the CNNs, the skin disease NN achieved an accuracy rate of 91\% (\textpm 0.2\%), with a cross entropy of 0.6. When the symptoms or images were fed into the device, the softmax function generated probabilities of certainty for different conditions. The features of the CNN were identified through t-distributed Stochastic Neighbor Embedding (t-SNE) to visualize the high-dimensional data of the last hidden layer representation of skin images. Each point represents a single dermoscopic image, and t-SNE allows for the visualization of how the CNN clusters similar image classes (see Fig. 6). 
\begin{figure}[htbp]
\centerline{\includegraphics{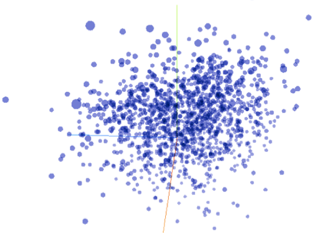}}
\caption{Output of t-SNE Algorithm in Three Dimensions}
\label{fig}
\end{figure}

\section{Further Research}
Doctors use contextual factors such as medical, family, and social history; medical test results; and many more, to conclusively make a successful diagnosis. In the future, adding these inputs to the neural network structure could result in a helpful tool by increasing the accuracy of diagnosis. By increasing the size of the dataset further, utilizing crowd-sourced data, the algorithm will become smarter and more accurate at diagnoses over time. The device will soon be tested out in the field with board-certified doctors to get their view on how the algorithm or implementation could be improved. Possibly utilizing natural language processing, the algorithm soon could process normal speech and provide a diagnosis. Increasing the options in the treatment list would also increase the usefulness of the device, as the database utilized did not have all possible treatments for the diseases that are diagnosable through the device. In order to provide more quantitative data for the DNN, sensors for biometric readings (heart rate, blood pressure, temperature, etc.) could be implemented into the device to provide immediate analysis. As the CNN and DNN become more complex, a more computationally powerful (albeit more costly) processor such as the NVIDIA Jetson, would become necessary for offline, on-edge analysis. Generally, in the future, the device will be refined by making it easier to use and more accurate and efficient. 

\section{Conclusion}
A singular device was created that is able to diagnose medical diseases. The basic device costs \$20 and can diagnose diseases entirely offline, with an accuracy rate that is comparable to modern doctors (80\%) \cite{b14}. The device achieved all the engineering goals outlined and far exceeds the current standard for automatic diagnosis. It does not require internet access, is accurate and cheap, and the code is free and public. This scalable method for diagnosis holds the potential for clinical impact around the world. It has the possibility of assisting in streamlining medical facilities and helping healthcare personnel. While it is understood that doctors use other contextual factors during diagnosis, other than raw symptoms and visual evidence, the ability to allow diagnosis comparable to medical professionals individually, has the power to profoundly expand access to vital medical care. \\ \\ \\

\end{document}